\documentclass[11pt,a4paper]{article} 
\usepackage{jcappub}
\usepackage{framed}

\newcommand{\bea}{\begin{eqnarray}}
\newcommand{\eea}{\end{eqnarray}}
\newcommand{\pd}{\partial}
\newcommand{\nd}{\nabla}
\newcommand{\nn}{\nonumber\\}
\newcommand{\be}{\begin{equation}}
\newcommand{\ee}{\end{equation}}
\newcommand{\mc}{\mathcal}
\newcommand{\tx}{\mathrm}

\title{A no-hair theorem for stars in Horndeski theories}

\author[a]{A.~Leh\'ebel,}

\author[a]{E.~Babichev,} 

\author[a]{C.~Charmousis}

\affiliation[a]{Laboratoire de Physique Th\'eorique, CNRS, Univ. Paris-Sud, \\ Universit\'e Paris-Saclay, 91405 Orsay, France}

\emailAdd{eugeny.babichev@th.u-psud.fr}
\emailAdd{christos.charmousis@th.u-psud.fr}
\emailAdd{antoine.lehebel@th.u-psud.fr}

\abstract{We consider a generic scalar-tensor theory involving a shift-symmetric scalar field and minimally coupled matter fields. We prove that the Noether current associated with shift-symmetry vanishes in regular, spherically symmetric and static spacetimes. We use this fact to prove the absence of scalar hair for spherically symmetric and static stars in Horndeski and beyond theories. We carefully detail the validity of this no-hair theorem.}

\begin{document}

\maketitle

\section{Introduction}

Compact astrophysical objects, like neutron stars or black holes, are most likely to open interesting and enlightening observational windows on gravity in the forthcoming years. With the recent gravitational wave detection of binary black hole mergers~\cite{Abbott:2016blz,Abbott:2017vtc}, it is most probably a matter of time before we  detect neutron star mergers. Such forthcoming observations will scrutinize the strong field structure of these objects, determining  whether their dynamics obey General Relativity (GR) or deviate from its predictions. So far, GR accounts for all local gravitational phenomena, from table-top experiments to solar system tests. Remarkably, GR also fits very well with black hole merger tests mentioned above~\cite{Abbott:2016blz,Abbott:2017vtc}. If the validity of GR is to be questioned, it is rather at very low curvature scales or at very large distances, some $10^{15}$ greater than local scales, on the level of cosmology. Indeed, the accelerated expansion of the universe requires the introduction of an enigmatic energy component, designated as dark energy. Quantum field theory provides a qualitative candidate for dark energy, a cosmological constant, unfortunately predicting an energy density at least 54 orders of magnitude greater than the observed one~\cite{Martin:2012bt}. Even at galactic scales, GR together with the Standard Model of particle physics fails to account for about 80\% of the gravitationally interacting matter. This suggests that GR is missing a vital component in the deep infrared sector, inviting one to consider modified gravity theories.

An important category amongst alternative gravity theories are scalar-tensor theories. In this framework, a scalar field mediates an additional long-range interaction, which can provide an alternative answer to some of the dark sector puzzles. However, if one desires a consistent modification of gravity, one should recover the observed GR predictions at local scales. A good alternative theory should therefore exhibit a so-called `screening mechanism' of the additional gravitational degrees of freedom, such as the Vainshtein mechanism~\cite{Vainshtein:1972sx,Babichev:2013usa}. 
This should be true, at the level of local distance scales, both in weak and strong gravity regimes where GR is put to the test by observations. Although this screening is quite well understood for weak gravity scales, the
situation is far more complex for strong gravity regimes, for example in the case of
solitary or binary neutron stars and black holes. 

How different from GR can scalar-tensor strong field solutions be? Should we expect discrepancies at the level of the background solution, or rather in the gravitational wave emission? Or again in none of these sectors? In GR, there exist uniqueness theorems concerning black holes, stating that the Kerr-Newman family covers the whole space of physically admissible black hole solutions. These theorems are not easily translated to scalar-tensor theories. Usually, more restrictive assumptions must be added to maintain their validity (see~\cite{Sotiriou:2015pka} for a review about the scalar-tensor case). For example, in the case of static black holes, there exists a no-hair theorem \cite{Hui:2012qt}, which under specific hypotheses \cite{Sotiriou:2014pfa,Babichev:2016rlq}, dictates that black holes are identical to GR with a constant scalar field. If the no-hair hypotheses are not met, the theorem provides quite stringent guidelines on how to obtain non-trivial scalar-tensor black holes, and some solutions have been found 
\cite{Sotiriou:2014pfa, Rinaldi:2012vy, Anabalon:2013oea, Minamitsuji:2013ura, Babichev:2013cya, Babichev:2015rva, Charmousis:2015aya, Charmousis:2014zaa, Babichev:2017guv}. 
They either deviate from the Schwarzschild metric, or on the contrary are essentially indistinguishable from GR solutions at the background level. Little is known on stationary solutions \cite{Maselli:2015yva,Babichev:2017guv} and even less on binaries. On the other hand, when considering star solutions, the uniqueness theorems break down even in GR. Would the introduction of a fundamental scalar field enrich the spectrum of GR star solutions, and in what way?

In this paper, we provide the first step, namely a no-hair theorem, investigating the spherically symmetric and static star configurations in scalar-tensor theories with minimal matter coupling. We show that these configurations are generically trivial under the key requirement that only derivatives of the scalar field are present in the action (with some extra technical assumptions). We prove in Sec.~\ref{sec:Jr} that in a regular, spherically symmetric and static spacetime, there cannot be a non-vanishing and time-independent scalar flow. In Sec.~\ref{sec:nohair}, we use this fact in the framework of Horndeski and beyond theories. Horndeski theory is the most general theory involving a metric and a scalar field while maintaining second-order field equations \cite{Horndeski:1974wa}. Requiring second-order equations is enough to ensure the absence of an Ostrogradski instability. Recently though, it was pointed out that one can consistently go beyond this requirement~\cite{Gleyzes:2014dya,Gleyzes:2014qga,Zumalacarregui:2013pma}. Indeed, certain theories lead to third-order equations of motion, but at the same time they keep a dynamical structure with the right number of canonical degrees of freedom \cite{,Langlois:2015cwa,Deffayet:2015qwa,Langlois:2015skt,Crisostomi:2016tcp,Crisostomi:2016czh}. Such degenerate theories go by the name of beyond Horndeski theories, and we investigate them as well in this paper. We examine in detail the various assumptions that lead us to the no-hair result, and enumerate the possible ways out. We make the link with black hole configurations, underlining the similarities and differences with stars. Finally, Sec.~\ref{sec:conc} contains our conclusions.

\section{No influx on stars}
\label{sec:Jr}

We consider a metric theory that involves a non-minimally coupled scalar field, entering the action only via its derivatives\footnote{A calculation similar to the one presented in this section was first carried out in \cite{Barausse:2017gip}.}. We allow the presence of standard matter, minimally coupled to the metric. Namely, we analyze the following action:
\bea
S= \displaystyle\int{\tx{d}^4x \sqrt{-g}\: \mc{L}_\tx{g}\left[g_{\mu\nu}, g_{\mu\nu,\rho_1},...,g_{\mu\nu,\rho_1...\rho_p};\phi_{,\rho_1},...,\phi_{,\rho_1...\rho_q}\right]}+S_\tx{m}\left[g_{\mu\nu}; \Psi\right],
\label{eq:action}
\eea
with $p,q\geq 2$ and finite. $\phi$ denotes the non-standard scalar field; the matter action $S_\tx{m}$ involves matter fields collectively denoted as $\Psi$. The latter obey the weak equivalence principle. By construction, the action is not affected by a homogeneous shift of the scalar field $\phi \rightarrow \phi+C$. There accordingly exists a conserved Noether current:
\bea
\label{eq:covcur}
J^\mu=\frac{1}{\sqrt{-g}}\, \frac{\delta S[\phi]}{\delta (\partial_\mu\phi)}.
\eea
We restrict our attention to a spherically symmetric and static geometry which is regular everywhere.
Black holes are thus excluded from our analysis; rather, we have in mind star configurations. The scalar field is assumed to respect the symmetries of spacetime\footnote{The staticity of the scalar is a choice we make. Time-dependence of the scalar is known to give well-behaved black holes and star solutions~\cite{Babichev:2013cya, Cisterna:2015yla, Charmousis:2014mia} for some shift-symmetric theories among (\ref{eq:action}).}. The general ansatz that respects these conditions reads:
\bea
\tx{d} s^2 &=& -h(r) \tx{d} t^2 + \dfrac{\tx{d} r^2}{f(r)} + r^2 (\tx{d}\theta^2 + \sin^2 \theta \tx{d}\phi^2),
\\
\phi &=& \phi(r),
\label{eq:ansatz}
\eea
with $f$ and $h$ two arbitrary functions of the radial coordinate. Such an ansatz for the scalar field and the metric imposes that the only non-vanishing component of the Noether current is the radial one. The conservation equation is then easy to integrate. It reads:
\bea
\nd_\mu J^\mu = \dfrac{\tx{d}}{\tx{d}r}\left(\sqrt{\dfrac{h}{f}} r^2 J^r\right) = 0,
\eea
the generic solution of which is:
\be
J^r= -\dfrac{Q}{r^2} \sqrt{\dfrac{f}{h}},
\ee
where $Q$ is a free integration constant. One can think of $Q$ as a scalar charge, in the sense that it gives the far-away behavior of the scalar field. For instance, assuming flat asymptotics, the presence of a standard kinetic term $(\pd\phi)^2$ in the action imposes that $\phi$ will decay like $Q/r$ at infinity (up to an arbitrary constant).

\begin{figure}[ht]
\begin{center}
\includegraphics[width=.5\textwidth]{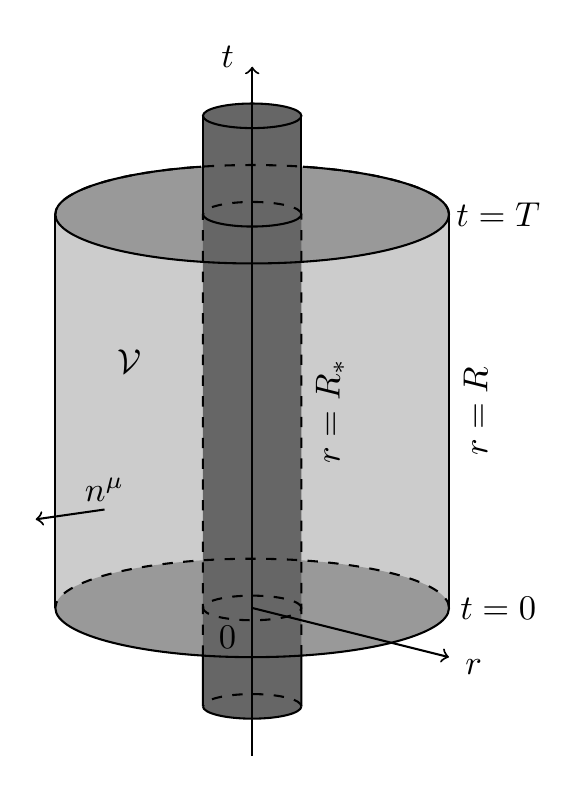}
\caption{Schematic representation of the spacetime. The vertical axis corresponds to time, and only two spatial dimensions are represented transversally. The light gray cylinder represents the 4-volume $\mathcal{V}$ over which we integrate. The dark gray cylinder of radius $R_\ast$ is the worldtube of some star, located around the origin. Intermediate gray surfaces are constant time slices. $n^\mu$ is the outward-pointing unit vector normal to the boundary of $\mathcal{V}$.}
\label{fig:tubes}
\end{center}
\end{figure}

In a static configuration, it seems problematic that a non-vanishing flux can indefinitely flow towards the origin of coordinates. Indeed, we will now prove that such star configurations are forbidden, i.e. that $Q$ has to vanish. To draw a parallel, Maxwell's equation in vacuum $\tx{div}(\textbf{E})=0$ also locally allows a radial electric field $\textbf{E} = C \textbf{r}/r^3$. However, integration over an extended domain imposes that $\textbf{E}$ actually vanishes in the absence of a charge distribution. Similarly here, $J^\mu$ will vanish because it is not sourced. To prove that $J^\mu$ is indeed zero, we integrate the conservation equation over a particular 4-volume $\mathcal{V}$. This volume is defined as the interior of a 2-sphere of radius $R$ between time $t = 0$ and $T$, as displayed in Fig. \ref{fig:tubes}. Matter fields are not required to be located in a compact region. To set the ideas though, one can think of a star located at the origin of coordinates, with matter fields present below the surface $r = R_\ast$.

We have explicitly required that the geometry is regular everywhere, in particular at the origin of coordinates. Therefore, $\mathcal{V}$ is a compact manifold with boundary $\pd \mathcal{V}$. From Fig.~\ref{fig:tubes}, it is easy to understand that $\pd \mathcal{V}$ consists of the top, bottom and side of the light gray cylinder. More precisely, the top and bottom are the interior of 2-spheres of radius $R$, at time $T$ and 0 respectively; the side $\mathcal{S}$ is the Cartesian product of the 2-sphere of radius $R$ with the segment of time $[0; T]$. The Gauss-law version of Stokes's theorem for $J^\mu$ then reads:
\be
\displaystyle\int_\mathcal{V}{\nd_\mu J^\mu} = \displaystyle\oint_{\pd\mathcal{V}}{n_\mu J^\mu},
\ee
where $n^\mu$ is the outward-pointing unit vector normal to $\pd \mathcal{V}$. The left-hand side vanishes because of current conservation. The integral over the top and bottom of the cylinder vanishes because $n^\mu$ and $J^\mu$ are orthogonal on these surfaces. Thus,
\bea
0 &=& \displaystyle\int_\mathcal{S}{n_\mu J^\mu}
\nn
&=& \displaystyle\int_\mathcal{S}{\tx{d}t \:\tx{d}^2\Omega \:\dfrac{R^2}{\sqrt{f(R)}}\sqrt{h(R)} J^r(R)}
\nn
&=& 4 \pi T R^2 \sqrt{\dfrac{h(R)}{f(R)}} J^r(R).
\eea
This is valid for arbitrary $R$. The only solution is to set $J^r=0$, or equivalently $Q=0$. Therefore, a permanent influx of scalar current is forbidden for a star configuration. Note that, so far, our proof follows a path similar to the one developed in~\cite{Yagi:2015oca}. Reference~\cite{Yagi:2015oca} deals in particular with a special type of Lagrangian, where a linear coupling between the scalar field and the Gauss-Bonnet invariant is present. Such a theory is included in the class~(\ref{eq:action}) up to boundary terms, because the Gauss-Bonnet invariant is a total divergence. Due to this fact precisely, the associated scalar field equation can be put under the form of a conservation equation. Using this fact and under some assumptions on the faraway behavior of the metric, the authors of~\cite{Yagi:2015oca} prove that the scalar field cannot exhibit a $1/r$ decay at infinity. In the next section, we show that shift symmetry together with certain hypotheses---notably excluding the cause of linear coupling to Gauss-Bonnet invariant--- allows to establish a considerably stronger result for the very general theory~(\ref{eq:action}).

\section{No scalar field around stars}
\label{sec:nohair}

In this section, we specialize to Horndeski and beyond theories. Using the result of the previous section, we prove a no-hair theorem for stars. The shift-symmetric version of Horndeski and beyond theories belongs to the class~(\ref{eq:action}), so the above discussion is valid. Explicitly, shift-symmetric Horndeski and beyond theories are labelled by 6 arbitrary functions $G_2$, $G_3$, $G_4$, $G_5$, $F_4$ and $F_5$ of the kinetic density $X=-\pd_\mu\phi\pd^\mu\phi/2$:
\bea
\label{action}
S &=& \int\, \sqrt{-g}\, \tx{d}^4x \left(\mc{L}_2+\mc{L}_3+\mc{L}_4+\mc{L}_5 +\mc{L}^{\rm bH}_4+\mc{L}^{\rm bH}_5 \right),
\eea
with
\bea
\label{eq:Haction}
\mc{L}_2 &=& G_2(X) ,
\\
\mc{L}_3 &=& -G_3(X) \Box \phi ,
\\
\mc{L}_4 &=& G_4(X) R + G_{4X} \left[ (\Box \phi)^2 -(\nabla_\mu\nabla_\nu\phi)^2 \right] ,
\\
\mc{L}_5 &=& G_5(X) G_{\mu\nu}\nabla^\mu \nabla^\nu \phi - \frac{1}{6} G_{5X} \big[ (\Box \phi)^3 - 3\Box \phi(\nabla_\mu\nabla_\nu\phi)^2 \nn
&~&~~ + 2(\nabla_\mu\nabla_\nu\phi)^3 \big],
\\
\mc{L}_4^{\tx{bH}} &=& F_4(X) \epsilon^{\mu \nu \rho \sigma} {\epsilon^{\alpha \beta \gamma}}_\sigma \nd_\mu\phi \nd_\alpha\phi \nd_\nu \nd_\beta\phi \nd_\rho \nd_\gamma\phi ,
\\
\mc{L}_5^{\tx{bH}} &=& F_5(X) \epsilon^{\mu \nu \rho \sigma} \epsilon^{\alpha \beta \gamma \delta} \nd_\mu\phi \nd_\alpha\phi \nd_\nu \nd_\beta\phi \nd_\rho \nd_\gamma\phi \nd_\sigma \nd_\delta\phi ,
\eea
where a subscript $X$ stands for the derivative with respect to $X$, $R$ is the Ricci scalar, $G_{\mu\nu}$ is the Einstein tensor, $(\nabla_\mu\nabla_\nu\phi)^2 = \nabla_\mu\nabla_\nu\phi \nabla^\nu\nabla^\mu\phi$ and $(\nabla_\mu\nabla_\nu\phi)^3=\nabla_\mu\nabla_\nu\phi \nabla^\nu\nabla^\rho\phi \nabla_\rho\nabla^\mu\phi$. The $F_4$ and $F_5$ functions correspond to beyond Horndeski terms. If one leaves them aside, one gets back to the original Horndeski theory. Using the ansatz~(\ref{eq:ansatz}), the radial component of the Noether current reads
\bea
\label{eq:Jrexpr}
J^r=&-&f \phi' G_{2X} - f\dfrac{r h' + 4 h}{r h} X G_{3X} + 2 f \phi' \dfrac{f h - h + r f h'}{r^2 h}G_{4X} + 4 f^2 \phi' \dfrac{h +rh' }{r^2h} X G_{4XX} 
\nn
&-&f h' \dfrac{1 - 3 f}{r^2h} X G_{5X} +2 \dfrac{h' f^2}{r^2h} X^2 G_{5XX}+ 8 f^2 \phi' \dfrac{h +rh' }{r^2h} X (2 F_4+X F_{4X})
\nn
&-& 12 \dfrac{f^2 h'}{rh} X^2 (5 F_5+2 X F_{5X})
,\eea
where a prime stands for a derivative with respect to the radial coordinate $r$. We know from the previous section that $J^r$ vanishes. To conclude about the trivial character of the scalar field, we also assume that the geometry is asymptotically flat and that the scalar field decays at infinity: $f,h \rightarrow 1$ and $\phi' \rightarrow 0$ as $r\rightarrow \infty$. Finally, we impose requirements on the form of the action. We first require the presence of a standard kinetic term $X \subseteq G_2$. Secondly, we ask that the action (\ref{eq:action}) is analytic around the point $\phi=constant$, so that one can expand it perturbatively around a trivial scalar configuration. This is achieved only when the $G_i$ and $F_i$ functions are analytic themselves.

Under these conditions, (\ref{eq:Jrexpr}) can be put under the form
\bea
J^r  = \phi' \mathcal{J}(\phi',f,h'/h,r),
\eea 
where $\mc{J}$ depends analytically on $\phi'$ since $X=-\phi'^2 f/2$ and the $G_i$ and $F_i$ functions are analytic. The vanishing of $J^r$ locally splits the solutions into two branches: at any point, either $\phi'$ or $\mc{J}$ vanishes. Let us demonstrate that, under the assumptions we made, $\phi'=0$ is the solution. Indeed, a careful examination of the various terms in~(\ref{eq:Jrexpr}) shows that $\mc{J}(0,1,0,r) = -1$. Therefore, using asymptotic flatness, $\mathcal{J}$ tends towards -1 at infinity. Let us consider a radius $r_1$ large enough for $\mathcal{J}$ to remain negative when $r\geq r_1$. Since $J^r$ cancels exactly everywhere, so does $\phi'$ in the whole outer region $r\geq r_1$. As a consequence, the contribution of the scalar to the field equations vanishes in this region, and the solution is uniquely given by GR. It is Schwarzschild geometry if no matter is present, or else some specific (GR) star solution. We must finally establish that $\phi'$ also cancels in the inner region\footnote{One could think that the system of field equations together with the GR-like initial conditions at $r=r_1$ constitute a well posed Cauchy problem, thus forcing the solution to be GR in the inner region. However, because of the branch structure, the Cauchy-Kowalevski theorem does not apply.}. For $r\geq r_1$, since $\phi'$ vanishes, the expression of $\mc{J}$ is easily derived from~(\ref{eq:Jrexpr}):
\bea
\mc{J}=-f+2G_{4X}(0)G^{rr}
\label{eq:curlyJ}
,\eea
where $G^{rr}$ is the $(rr)$ component of Einstein tensor. We see that the presence of a linear term in $G_4(X)$ is important. At this point, we make a last simplifying assumption: that $G_{4X}(0)=0$. The case $G_{4X}(0)\neq0$ is considered separately in Appendix~\ref{sec:John} where we argue that, even in this situation, one still cannot build any nontrivial viable solution.
Thus, as long as $\phi'=0$, $\mc{J}=-f$. When entering the region $r<r_1$, $\mc{J}$ remains non-zero by continuity. 
Accordingly, $\phi'$ does not deviate from 0. This in turns implies that $\mc{J}$ remains equal to $-f$. Since the geometry is regular, $f$ does not vanish anywhere, and neither does $\mc{J}$. We have no other choice than to stick to the $\phi'=0$ branch. We are left with GR as the unique solution over $0\leq r<+\infty$.

Let us summarize what we have proven, and under which conditions. Consider any shift-symmetric Horndeski or beyond theory with a minimally coupled matter sector, as in~(\ref{action}). We assume that:
\begin{enumerate}
\item the metric and scalar field are regular\footnote{By regular, we mean that the metric and the scalar field are $C^1$, and that the metric is invertible in the chosen system of coordinates.} everywhere, spherically symmetric and static,
\item spacetime is asymptotically flat with $\phi'\rightarrow 0$ as $r\rightarrow \infty$,
\item there is a canonical kinetic term $X$ in the action, and the action is analytic around a trivial scalar field configuration.
\end{enumerate} 
Under these hypotheses, we conclude that $\phi$ is constant, and that the only solutions are the GR ones. In particular, star solutions are identical to their GR counterpart.

The result and assumptions of this theorem are very close to the no-hair theorem for black holes formulated by Hui and Nicolis~\cite{Hui:2012qt}. Of course, the latter does not assume that spacetime is regular everywhere. Instead, it requires that the norm of the current $J^2$ is finite at any point, especially when approaching the horizon. Interestingly, this extra assumption imposes that $J^r=0$ everywhere. In the situation we examined in this paper, the regularity of spacetime unequivocally imposes that $J^r$ vanishes. When studying a black hole, one has to impose an additional regularity assumption on the current. Apart from this subtlety, the end of the proof is similar in both cases\footnote{In~\cite{Hui:2012qt}, the step $J^r=0\Rightarrow\phi'=0$ follows simply from Eq.~(\ref{eq:curlyJ}). Indeed, the absence of matter imposes that $G^{rr}=0$, so $\mc{J}$ cannot vanish down to the horizon, sparing one the discussion of Appendix~\ref{sec:John}.}, because it only relies on asymptotic flatness. In the case of black holes, asking for the norm of the current to be finite is not a trivial requirement. The question amounts to know whether a divergence of $J^2$ has any observable effect. 
In~\cite{Sotiriou:2014pfa}, black hole solutions were studied in the following theory:
\bea
S=\dfrac{M_\tx{Pl}^2}{16\pi}\displaystyle\int{\tx{d}^4x\sqrt{-g}\left(R-\pd_\mu\phi \pd^\mu\phi +\alpha\phi \hat{G}\right)},
\label{eq:GBaction}
\eea
where $M_\tx{Pl}$ is the Planck mass, $\alpha$ is some dimensionful coupling constant and $\hat{G}=R_{\mu \nu \rho \sigma}R^{\mu \nu \rho \sigma} - 4 R_{\mu \nu} R^{\mu \nu} +R^2$ is the Gauss-Bonnet invariant. This theory belongs to the class~(\ref{eq:action}) because the last term is equivalent to the choice~$G_5(X)=-4\alpha \ln|X|$. Although not stated in~\cite{Sotiriou:2014pfa}, the norm of the current diverges at the horizon for the exhibited solutions, see~\cite{Babichev:2016rlq}. However, the metric is regular everywhere. Furthermore, actions of type~(\ref{eq:action}) do not couple the scalar current to matter by construction. It is therefore not obvious whether such solutions should be ruled out or not. The answer to this question necessarily involves the study of the dynamical collapse of stars into black holes in this theory\footnote{For a test scalar field, a study was carried out in \cite{Benkel:2016rlz}.}. For the moment, nothing guarantees that the endpoint of such a collapse would be the solution described in~\cite{Sotiriou:2014pfa}. In particular, the solution requires a very particular tuning between the mass and scalar charge of the black hole.

A number of ways out exist for the black hole no-hair theorem in Horndeski and beyond theories (see~\cite{Babichev:2016rlq} for a recent review). Since the theorem we propose relies on very similar assumptions, the same possibilities are offered to escape from it. In particular, one should be able to find hairy star solutions embedded in an asymptotically de Sitter universe. Some ways out have actually already been explored: a series of papers~\cite{Cisterna:2015yla,Cisterna:2016vdx,Maselli:2016gxk} focused on solutions where the scalar field is allowed to acquire time-dependence, and also on static solutions with no canonical kinetic term. Finally, one could circumvent the theorem by constructing star solutions with non-analytic $G_i$ or $F_i$ functions, as was done for black holes in~\cite{Babichev:2017guv}. The solution presented in~\cite{Sotiriou:2014pfa} clearly cannot be used for describing the exterior of a star, since it possesses a non-vanishing current. Rather, the faraway behavior of the star solutions of theory~(\ref{eq:GBaction}) is the one derived in~\cite{Babichev:2017guv}, with a scalar decaying as $\mathcal{O}(1/r^4)$, and $\mathcal{O}(1/r^7)$ corrections to the Schwarzschild metric{\footnote{In this case, $J^r=0$, but the no-hair theorem presented here does not hold since $G_5 =-4 \alpha\ln|X|$ is not analytic at $X=0$.}}.

\section{Conclusion}
\label{sec:conc}

We proved that a shift-symmetric scalar-tensor theory cannot accommodate a non-vanishing Noether current as soon as spacetime is asked to be regular, static and spherically symmetric. Using this result, some complementary assumptions allow us to conclude that stars cannot develop scalar hair in Horndeski and beyond theories. Star solutions are therefore identical to GR solutions for these theories. This extends what was known previously in the case of black hole geometries. Most of the ways out known for black holes still allow to evade the theorem for stars, with the notable exception of black hole solutions with non-vanishing current~\cite{Sotiriou:2014pfa}. An unexplored class of hairy star solutions would be generated by generalizing the results of~\cite{Babichev:2017guv} to star configurations.

Interestingly, our result is complementary to what Barausse and Yagi derived in~\cite{Barausse:2015wia}. They investigated gravitational wave emission in shift-symmetric Horndeski theory. Under assumptions that are very similar to those we made here, they showed that, 
at leading Post-Newtonian order, gravitational wave emission for star binaries is not modified with respect to GR. This result concerns wave propagation, while our theorem focuses on the static structure of stars. However, they both combine to point at the absence of differences between GR and shift-symmetric (beyond) Horndeski theories. 

A natural question to ask is whether this no-hair theorem can be extended to stationary solutions. The result of~\cite{Barausse:2015wia} about gravitational waves only requires stationarity, which might be an indication that the theorem developed presently remains valid. Additionally, an extension of the theorem for stars would probably be transposable to black holes, since the arguments are very similar in both situations. This is an interesting track for future studies.

\acknowledgments
We thank Enrico Barausse for pointing out reference~\cite{Barausse:2017gip}. The authors acknowledge financial support from the research program, Programme national de cosmologie et galaxies of the CNRS/INSU, France and from the project DEFI InFIniTI 2017. EB was supported in part by Russian Foundation for Basic Research Grant No. RFBR 15-02-05038.

\appendix

\section{The case \texorpdfstring{$G_{4X}(0)\neq0$}{G4X(0) is non-zero}}
\label{sec:John}

Let us resume the discussion at Eq.~(\ref{eq:curlyJ}). We can put the $(rr)$ metric field equation in the following form:
\bea
G^{rr}=8\pi(T_\phi^{rr}+T_\tx{m}^{rr})
\label{eq:Err1}
,\eea
where $T_\phi^{\mu\nu}$ and $T_\tx{m}^{\mu\nu}$ are the stress-energy tensors that originate respectively from the scalar field and matter sector. In the outer region, and as long as we follow the branch $\phi'=0$, $T_\phi^{\mu\nu}$ identically vanishes. Therefore, noting $P= T_\tx{m}^{rr}/f$ and $\beta=G_{4X}(0)$, the function $\mc{J}$ takes the following form:
\bea
\mc{J}=f(16\pi \beta P-1)
.\eea
$P$ should be thought of as the pressure due to matter fields. As in Sec.~\ref{sec:nohair}, $f$ never vanishes because of regularity, and we must again follow the branch $\phi'=0$ as long as $\mc{J}\neq0$. However, it can now happen that the pressure reaches the critical value $1/(16\pi \beta)$. Let us assume this indeed happens at some radius $r=r_0$. Then, nothing a priori forbids the solution to jump from the first branch where $(\phi'=0,\mc{J}\neq0)$ to the second one, with $(\phi'\neq0,\mc{J}=0)$. In this case, the solution would differ from GR in the ball $r<r_0$. To go further, we need to take a closer look at the other field equations. For any Horndeski or beyond theory, the $(rr)$ metric field equation\footnote{To be precise, this is a judicious combination of Eq.~(\ref{eq:Err1}) together with the $J^r=0$ constraint, as in~\cite{Babichev:2016kdt}.} reads:
\bea
\label{eq:Err2}
&G_2& - \dfrac{2}{r^2h}(frh'+fh-h) G_4 + \dfrac{4f}{r^2h} (rh'+h) X G_{4X} - \dfrac{2}{r^2h}f^2h'\phi'XG_{5X}
\nn
&+&\dfrac{8f}{r^2h}(rh'+h)X^2F_4-\dfrac{24}{r^2h}f^2h'\phi'X^2F_5 = -M_\tx{Pl}^2 P.
\eea
Let us solve this equation in the vicinity of the possible transition from one branch to the other. The scalar field is assumed to be $C^1$, thus $\phi'$ is continuous at $r_0$ and remains small close to $r_0$. Thanks to the analyticity of the $G_i$ and $F_i$ functions, Eqs.~(\ref{eq:Err2}) and~(\ref{eq:Jrexpr}) can be put in the following form:
\bea
\dfrac{M_\tx{Pl}^2}{8\pi}\left(8\pi P-\dfrac{frh'+fh-h}{r^2h}\right)-\dfrac12 f\phi'^2\left(1-2\beta\dfrac{frh'+fh-h}{r^2h}+4\beta\dfrac{rh'+h}{r^2h}\right)&\phantom{lol}&
\nn
+\mc{O}\left[\phi'(r)^3\right]\underset{r\rightarrow r_0}{=} &0&,
\\
-1+2\beta \dfrac{fh-h+rfh'}{r^2h}+\mc{O}\left[\phi'(r)\right]\underset{r\rightarrow r_0}{=}&0&
,\eea
From these two equations, one can extract the behavior of $\phi'$ close to $r_0$. To do so, we notice that such a transition would happen inside the star and not at its surface, since the pressure $P$ would reach the value $1/(16\pi \beta)$ already. It is therefore natural to ask that $P'(r_0)\neq0$ (there may exists modified gravity models where $P'=0$  at some non-zero radius,  but there is absolutely no reason why this would happen precisely when $P=1/(16\pi \beta)$). We end up with
\bea
\phi' \underset{r\rightarrow r_0}{\sim} \sqrt{\dfrac{-M_\tx{Pl}^2 r_0^2 P_0'}{(2\beta+r_0^2)f_0}} \sqrt{r_0-r},
\eea
where $P_0'$ and $f_0$ are respectively the value of $P'$ and $f$ at $r_0$. However, this square root behavior is problematic. It leads to a jump in $\phi' \phi''$, which enters the $(tt)$ component of the stress-energy tensor associated with $\phi$. Therefore, either the metric or the matter fields must undergo a similar jump, which contradicts the assumption about their $C^1$ character. Thus, even when $G_{4X}(0)\neq0$, no physically sensible solution seem to exist apart from GR.

\bibliographystyle{unsrt}
\bibliography{biblio}
\end{document}